\documentclass[12pt]{article}
\usepackage{amsmath,amssymb,amsthm,color}
\usepackage{graphicx}
\usepackage{cite}
\usepackage{epsfig}
\usepackage{float}

\renewcommand{\baselinestretch}{1.66}
\begin{document}

\title {Geometry- and field-diversified electronic and optical properties in bilayer silicene \\ }

\author{
\small Thi-Nga Do$^{1 \dag}$, Po-Hsin Shih$^{2}$, Godfrey Gumbs$^{3,4}$, and Ming-Fa Lin$^{5,6 \ddag}$ \\
\small $^1$ Department of Physics, National Kaohsiung Normal University, \\
\small Kaohsiung, Taiwan \\
\small $^{2}$Department of Physics, National Cheng Kung University, Tainan, Taiwan 701 \\
\small $^3$ Department of Physics and Astronomy, Hunter College of the City University of New York, \\
\small 695 Park Avenue, New York, NY 10065, USA \\
\small $^4$Donostia International Physics Center (DIPC),\\
\small P de Manuel Lardizabal, 4, 20018 San Sebastian, Basque Country, Spain \\
\small $^{5}$Hierarchical Green-Energy Materials Research Center, \\
\small National Cheng Kung University, Tainan, Taiwan 701 \\
\small $^{6}$Quantum Topology Center, National Cheng Kung University, Tainan, Taiwan 701.}

\renewcommand{\baselinestretch}{1.66}
\maketitle

\renewcommand{\baselinestretch}{1.66}

\begin{abstract}

The generalized tight-binding model has been developed to thoroughly explore the essential electronic and optical properties of AB-bt bilayer silicene. They are greatly diversified  by the buckled structure, stacking configuration, intralayer and interlayer hopping integrals, spin-orbital couplings; electric  and magnetic fields (${E_z\hat z}$ $\&$ ${B_z\hat z}$).
There exist the linear, parabolic and constant-energy-loop dispersions, multi-valley band structure and semiconductor-metal transition as $E_z$ varies.
The $E_z$-dependent magnetic quantization exhibits the rich and unique Landau Levels (LLs) and magneto-optical spectra.
The LLs have the lower degeneracy, valley-created localization centers, unusual distributions of quantum numbers, well-behaved and abnormal energy spectra in $B_z$-dependences, and the absence of anti-crossing behavior.  A lot of pronounced magneto-absorption peaks occur at a very narrow frequency range, being attributed to diverse excitation categories. They have no specific selection rules except  that the Dirac-cone band structures are driven by the critical electric fields. The optical gaps are  reduced by  $E_z$, but enhanced by $B_z$, in which the threshold  channel might dramatically change in the formed case.
The above-mentioned characteristics are in sharp contrast with those of layered graphenes.

\vskip 1.0 truecm

\par\noindent $\dag$ Corresponding author: E-mail: sofia90vn@gmail.com  \\
\par\noindent $\ddag$ Corresponding author: Tel: +886-6-275-7575; Fax: +886-6-74-7995;\\
 E-mail: mflin@mail.ncku.edu.tw  \\

\end{abstract}

\pagebreak
\renewcommand{\baselinestretch}{2}
\newpage

\section{Introduction}
\label{sec:1}

Layered condensed matter systems, with varied physical properties and many potential device applications, have so far attracted a great deal of experimental and theoretical attention \cite{KIB;SSC146, TND;C2015, YKH;SR2014, JEP;JPCC119, JYW;PRB94, YB;PRB89, MD;NL15, PHS;SR2017, FFZ;NM14, RBC;SR2017, HL;ACSN8, GG;PRB96, JYW;X2018, PA;AM30, FR;S357, RBC;X2018}.
Recently, theoreticians have developed reliable models in order to explore basic physical
properties of newly discovered two-dimensional (2D) materials, especially for their electronic and optical
properties when external electric and magnetic fields are applied.
Few-layer 2D materials are the main  focus, mainly because of their
eclectic lattice symmetries, their nano-scaled thickness, and their inherently unique interactions.
Group-IV and -V 2D systems, which have been successfully fabricated in laboratory conditions, include graphene \cite{KIB;SSC146, TND;C2015, YKH;SR2014}, silicene \cite{JEP;JPCC119, JYW;PRB94, YB;PRB89}, germanene \cite{MD;NL15, PHS;SR2017}, tinene \cite{FFZ;NM14, RBC;SR2017}, phosphorene \cite{HL;ACSN8, GG;PRB96, JYW;X2018}, antimonene \cite{PA;AM30}, and bismuthene \cite{FR;S357, RBC;X2018}.
These systems are expected to play crucial roles in basic and applied sciences, in which the rich and unique properties are worthy of a systematic investigation. In the present work, we concentrate our  efforts to achieve a clear understanding of the optical absorption spectra of bilayer silicene, being  closely related to the electronic properties in the presence/absence of electric and magnetic fields. The generalized tight-binding, combined with the dynamic Kubo formula, are fostered for exploring the diversified essential properties thoroughly. All the intrinsic interactions and the external fields are taken into consideration simultaneously. The distinct behaviors of the energy bands, density of states (DOS), quantized Landau levels (LLs), spatial magneto-wave functions, and absorption spectra are discussed adequately.

\medskip
\par

Examination of the fundamental physical properties of 2D materials, including their electronic, optical, transport, and Coulomb excitations, is very helpful in justifying their importance in the field of  nanotechnology applications, such as the novel designs of nano-electronics, nano-optics, and energy storages \cite{A1, A2, A3, A4, A5, A6, A7}. The main characteristics of the electronic properties and optical spectra are available in the polarized and hyperspectral imaging for target identifications, as well as in the ultrafast light-intensity modulations of space-laser transmission. In particular, the unique optical excitations can be employed in the design of an  electro-optic sensor system which is easy to transport and assemble. Additionally, the valley-, orbital-, and spin-dependent LLs under a magnetic field provide effective channels for controlling angular momenta and spin projections of Fermion electrons. The spatial distribution of two electron spins when being transported by either spin-orbital coupling (SOC) or random impurity scatterings constitutes a basis for modern spintronics and valleytronics. Such characteristics are taken into account  in the design of the next-generation ultrafast transistors for on-chip image processing in photo-detections.  The novel phenomena in Coulomb excitations are utilized to design easily transportable, compact, low-power and reconfigurable devices in security and wideband optical communications.

\medskip
\par

An important ingredient in our calculations is the method for diagonalizing the Hamiltonians of emergent layered materials.
We have developed the generalized tight-binding model, which is based on sub-envelope functions of different sublattices. We also employ the dynamic Kubo formula in linear response theory. This procedure is capable of including all critical ingredients simultaneously, including the single- or multi-orbital chemical bondings, SOCs, magnetic and electric fields, uniform/modulated external fields, intralayer and interlayer hopping integrals, arbitrary numbers of layers, various stacking configurations, planar/curved surfaces, and hybridized structures
\cite{CY;IOP2017, TND;Si}.
The high degeneracy of each LL causes the numerical simulations to become very difficult when diagonalizing a large magnetic Hamiltonian. This difficulty may be overcome by using a band-like matrix.
This described procedure can considerably diversify the electronic and optical properties in calculating the band structures, valley and spin degeneracies, energy gap/band overlap, Van Hove singularities of the DOS, LL crossings/anti-crossings, magneto-optical selection rules, and diverse absorption structures.

\medskip
\par

Previous investigations have demonstrated that 2D group-IV structures exhibit particularly exceptional properties. In this regard, graphene has an sp$^{2}$-bonding planar structure, whereas silicene and germanene have buckled structures with a slightly mixed sp$^2$-sp$^3$ chemical bonding \cite{TND;C2015, JYW;PRB94, PHS;SR2017}. Additionally, the SOC is appreciable in the low-energy electronic properties of these two materials. The massless Dirac fermions of monolayer graphene primarily come from a honeycomb lattice with an underlying geometric symmetry. This  yields a gapless semiconductor whose DOS vanishes at the Fermi level. Silicene and germanene are semiconductors and possess direct  band gaps, in which the Dirac cones are distorted and separated
by the significant SOCs. All the group-IV 2D systems have both valley and spin degeneracies.
Central to the manipulation of energy bands is to engineer an energy gap which yields a material
for semiconductor applications.  This serves to diversify tailor-made electronic properties which we could
efficiently utilize.

\medskip
\par

\begin{figure}[H]
\centering
{\includegraphics[width=1.0\linewidth]{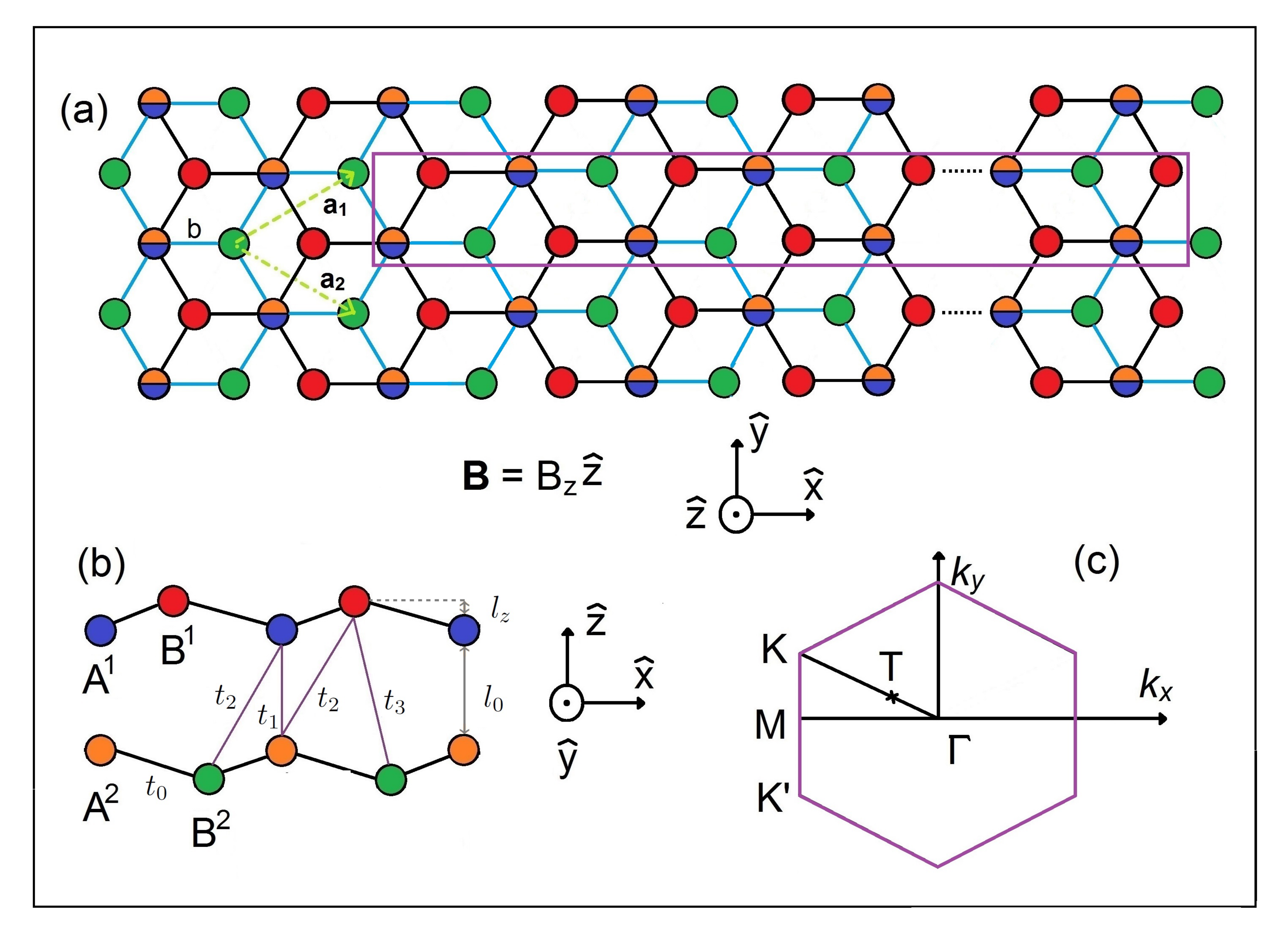}}
\caption{(Color online)  The geometric structure of AB-bt bilayer silicene with the top view (a) and side view (b). The enlarged unit cell under a uniform perpendicular
magnetic field is marked by the purple rectangular in (a). The intra- and inter-layer atomic interactions are presented in (b). The first Brillouin zone with the highly symmetric $\textbf{K}$ ($\bf{K^{\prime}}$) and $\bf{\Gamma}$ points and an extreme one, $\textbf{T}$, is shown in (c).}
\label{Figure 1}
\end{figure}

Up to this point in time, few-layer silicene has been successfully synthesized on different substrates \cite{PV;PRL2012, TL;NN2015}. Specifically, the AB-bt and AB-bb (bottom-bottom) configurations are observed by high-angle annular dark field scanning transmission electron microscopy \cite{RY;NC2016}. Also, AB-bt stacking with the lowest ground state energy among bilayer silicene systems is herewith chosen for a thorough investigation of its electronic and optical properties.
Its atomic structure with intra- and inter-layer atomic interactions and buckling order are depicted in Fig. \ref{Figure 1}, for which the first Brillouin zone is the same as that of layered graphene. With the significant SOCs, an electric field can destroy the $(x,y)$-plane mirror symmetry and thus lead to valley- and spin-split states in the AB-bt configuration. These critical factors dominate the low-energy physical properties. It is well known that a uniform perpendicular magnetic field creates highly degenerate LLs by quantization of the neighboring electronic states. The well-behaved LLs possess symmetric/antisymmetric spatial distributions in a localized range and their energy spectra display band-induced field dependencies. There also exist some perturbed/undefined LLs with frequent anti-crossing behaviors.
Specifically, the generalized tight-binding model can be used in conjunction with many-body and single-particle
theories when the eigenstates are represented in terms of sublattice envelope functions. The unified models are appropriate
for investigating a variety of properties including electronic properties, optical conductivities, quantum Hall effect, and Coulomb excitations.

\medskip
\par

\section{The generalized tight-binding model and Kubo formula}
\label{sec:2}

We develop the generalized tight-binding model to include all the critical factors in exploring the diverse electronic and optical properties of AB-bt bilayer silicene. The intra- and inter-layer atomic interactions, layer-dependent traditional and Bychkov-Rashba SOCs, and external electric and magnetic fields are taken into consideration simultaneously.
As to the bottom-top configuration, the first and second layers present the opposite buckled ordering, as shown in Fig. \ref{Figure 1}(b).
As a result, the vertical interlayer atomic interaction and SOCs are significantly strong. These dominate the low-lying electronic properties and absorption spectra. There are four atoms per unit cell, similar to that of bilayer graphene. On the other hand, silicene has a lattice constant of $a =3.86\,$\AA, which is much larger that that of graphene. Moreover, the buckled structure in silicene leads to an angle of $\theta =$ 78.3 $^{\circ}$ between the Si-Si bond and $z-$axis.
For AB-bt stacking, four sublattices of $A^{1,2}$ and $B^{1,2}$ lie on four distinct planes with intra- and inter-layer distances $l_z =0.46\,$\AA and $l_0 =2.54\,$\AA, respectively (Fig. 1(b)).
The above-mentioned intrinsic factors are responsible for the greatly diversified essential properties.

\medskip
\par

The tight-binding Hamiltonian may be written as \cite{TND;Si}

\begin{eqnarray}
\nonumber
H & = &\sum_{m,l,\alpha}(\epsilon_m^l+U_m^l)c_{m \alpha}^{\dagger l}c_{m \alpha}^{l}
+\sum_{ m,j  , \alpha, l, l^{\prime}} t_{mj}^{ll^{\prime}} c_{m \alpha}^{\dagger l}
c_{j \alpha}^{l^{\prime}}\\
\nonumber
&+&  \frac {i} {3\sqrt{3}} \sum_{ \langle \langle m,j \rangle \rangle, \alpha, \beta, l}
\lambda^{SOC}_{l} \gamma_l v_{mj} c_{m\alpha}^{\dagger l} \sigma_{\alpha\beta}^{z} c_{j\beta}^{l}\\
&-& \frac{2i}{3} \sum_{\langle\langle m,j \rangle \rangle, \alpha, \beta, l} \lambda^{R}_{l} \gamma_l u_{mj}
c_{m\alpha}^{\dagger l} (\vec{\sigma} \times \hat{d}_{mj})_{\alpha\beta}^{z} c_{j\beta}^{l}\ .
\end{eqnarray}

In this notation, $c_{m\alpha}^{l}$ ($c_{m\alpha}^{\dagger l}$) is the annilation (creation) operator, which destroys (creates) an electronic state at the $m$-th site of the $l$-th layer with spin polarization $\alpha$. The site energy, $\epsilon_m^l (A^l,B^l)$, arising from the chemical environment difference between the $A^l$ and $B^l$ sublattices, are defined as $\epsilon_m^l (A^l)$ = 0 and $\epsilon_m^l (B^l) = -0.12\,$eV. The height-induced Coulomb potential energy, $U_m^{l} (A^l,B^l)$, comes from the applied electric field.
The intra- and inter-layer hopping integrals, $t_{mj}^{ll^{\prime}}$, are related to the neighboring atomic interactions. The former (${t_0=1.13}\,$eV) and the latter (${t_1=-2.2}\,$eV, ${t_2=0.1}\,$eV; ${t_3=0.54}\,$eV) are clearly illustrated in Fig. 1(b). Especially, AB-bt bilayer silicene possesses the layer-dependent significant SOCs (the third and fouth terms), mainly owing to the very strong orbital hybridizations induced by the large inter-layer vertical hopping integral. They are optimized (${\lambda_1^{SOC}=0.06}\,$eV, ${\lambda_2^{SOC}=0.046}\,$eV, ${\lambda_1^{R}=-0.054}\,$eV; ${\lambda_2^{R}=-0.043}\,$eV) in order to reproduce the low-lying energy bands calculated by the first-principles method \cite{XW;PCCP2017, TND;Si}.

\medskip
\par

If AB-bt bilayer silicene is subjected to a uniform perpendicular magnetic field, Hamiltonian becomes a huge Hermitian matrix.
The enlarged unit cell due the vector potential (details in \cite{TND;Si}) is demonstrated in Fig. \ref{Figure 1}(a), e.g.,  10$^4$ silicon atoms under $B_z=40$ T.
For such a complex system, there is much difficulty in solving the eigenvalues and eigenstates.
We need to employ the band-like method and the spatial localizations of the magnetic wavefunctions to efficiently solve the LL eigenvalues and eigenfunctions. Moreover, based on the $B_z$-dependent evolution of LL energies, we can predict the main characteristics of LLs at the laboratory-produced field from those at higher field.
Each LL wavefunction, with the quantum number n, is given by

\begin{eqnarray}
\Psi (n,\mathbf{k})=\sum_{l=1,2}\sum_{m=1}^{R_B}\sum_{\alpha, \beta} [ A_{\alpha, \beta}^{l,m} (n,\mathbf{k}) |\psi_{\alpha, \beta}^{l,m} (A) \rangle + B_{\alpha, \beta}^{l,m} (n,\mathbf{k}) |\psi_{\alpha, \beta}^{l,m} (B) \rangle
  ].
\end{eqnarray}

We have $\psi_{\alpha, \beta}^{l,m}$ as the tight-binding function and $A_{\alpha, \beta}^{l,m} (n,\mathbf{k})$ ($B_{\alpha, \beta}^{l,m} (n,\mathbf{k})$) denotes the amplitude of the spatial distribution. They are strongly dependent on the lattice sites.
It should be noted that we have built up the tight-binding model using the $\vec{\bf{r}}$-scheme. That is, the calculations are based on the sublattices in an enlarged unit cell in coordinate space.
Moreover, the magnetic distribution width is much larger than the lattice constant, the amplitudes (the sub-envelope functions) in an enlarged unit cell could thus be considered as spatial distributions of the magnetic wave functions on the distinct sublattices. The LL wavefuntions provide sufficient information for us to explore the main phenomena of magnetic quantization and the optical selection rules. This is in sharp contrast with the method using the $\vec{\bf{k}}$-scheme as done by other research groups \cite{DR;PRB1976,  MK;AP1985}.

\medskip
\par

When AB-bt bilayer silicene is present in an electromagnetic field, electrons are vertically excited from occupied states to unoccupied ones with the same wave vectors.  The generalized tight-binding model combined with the dynamic Kubo formula is suitable for thoroughly exploring the optical excitations in the presence/absence of external fields. The zero-temperature spectral function related to the optical conductivity ($A(\omega) \propto \omega \sigma(\omega)$) is

\begin{eqnarray}
A(\omega) \propto
\sum_{c,v,m,m'} \int_{1stBZ} \frac {d\mathbf{k}}{(2\pi)^2}
 \Big| \Big\langle \Psi^{c} (\mathbf{k},m')
 \Big| \frac{   \hat{\mathbf{E}}\cdot \mathbf{P}   } {m_e}
 \Big| \Psi^{v}(\mathbf{k},m)    \Big\rangle \Big|^2 \nonumber
\end{eqnarray}
\begin{eqnarray}
 \times
Im \Big[      \frac{1}
{E^c (\mathbf{k},m')-E^v (\mathbf{k},m)-\omega - i\Gamma}           \Big].
\end{eqnarray}

The absorption function is associated with the velocity matrix element (the first term in Eq. (3)) and the joint density of states (JDOS; the second term).
The former, which is evaluated from the gradient approximation, as successfully utilized in layered graphene systems \cite{FM;PRB1994}, can determine the existence of vertical excitations/inter-LL transitions. That is, this dipole moment dominates the magneto-optical selection rules from the symmetries of the spatial distributions in the initial and final states. The latter reveals the van Hove singularities as the special absorption structures.
The generalized tight-binding model, combined with the Kubo formula, is useful for the thorough investigation of essential physical properties in many 2D materials, such as graphene \cite{CY;IOP2017}, silicene \cite{TND;Si}, tinene \cite{RBC;SR2017}, phosphorene \cite{JYW;X2018}, bismuthene \cite{RBC;X2018}, and others.

\section{Discussions on Calculated Results}
\label{sec:3}

\begin{figure}[H]
\centering
{\includegraphics[width=1.0\linewidth]{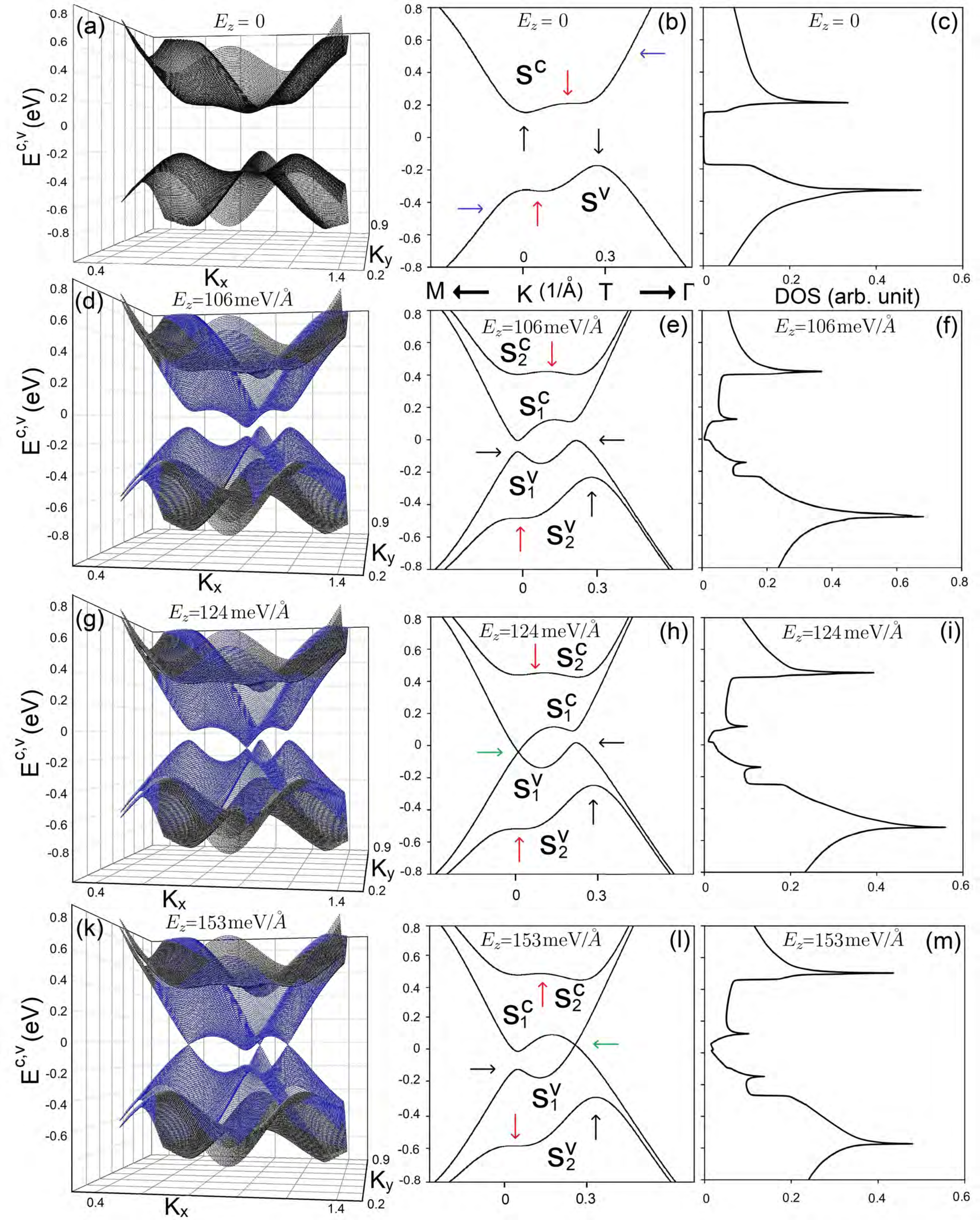}}
\caption{(Color online)  The 3D band structures, 2D energy bands along the high symmetry points and density of states (a)-(c) for ${E_z=0}$, (d)-(f) ${E_z=106}$ meV/$\text{\AA}$, (g)-(i) ${E_z=124}$ meV/$\text{\AA}$, and
(k)-(m) ${E_z=153}$ meV/$\text{\AA}$.}
\label{Figure 2}
\end{figure}

The unusual electronic structure in bilayer silicene, being characterized by the silicon $3p_{z}$ orbitals for the low-lying bands, covers the extremal band-edge states (black arrows), the partially flat energy dispersions along the specific direction (red arrows), the constant-energy loops (blue arrows), and the linear Dirac-cone structures (green arrows).
The second and third types could be regarded as 1D parabolic dispersions, revealing similar peak structures in the DOS, which we discuss below. The zero-field result is in agreement with that obtained from the first-principles method \cite{XW;PCCP2017}. The valence and conduction bands are very sensitive to an external electric field.
They are split into two pairs of spin-up- and spin-down-dominated energy subbands ($S^{c}_{1,2}$, $S^{v}_{1,2}$), as clearly indicated in Figs. \ \ref{Figure 2}(b), \ref{Figure 2}(e), \ref{Figure 2}(h), and \ref{Figure 2}(k). Spin degeneracy is absent in the presence of $E_z$ because of the elimination of mirror symmetry. With increasing $E_z$, the low-lying pair of energy subbands ($S^{c}_{1}$, $S^{v}_{1}$) gradually approach each other until the band gap is completely closed at the first critical field of $E_{z} \approx$ 106 meV/\AA, as is evident in Figs. \ref{Figure 2}(d)-\ref{Figure 2}(e), where the highest occupied valence state and the lowest unoccupied one have different wave vectors. On the other hand, the outer pair of ($S^{c}_{2}$, $S^{v}_{2}$) energy subbands monotonically moves away from the Fermi level. Beyond the critical electric field, the valence and conduction bands exhibit a slight overlap, and the linear Dirac cone structures are formed at the $\bf{K}$ ($\bf{K}$$^{\prime}$) point for the second critical electric field of $E_z=124$ meV/$\text{\AA}$ (Figs. \ref{Figure 2}(g)-\ref{Figure 2}(h)) and at the $\bf{T}$ point for the third one of $E_z=153$ meV/$\text{\AA}$ (\ref{Figure 2}(k)-\ref{Figure 2}(l))).
The Dirac points change from the occupied valence states into the unoccupied conduction ones in the further increase of $E_z$.

\medskip
\par

The density-of-states directly reflects the main features of the unusual energy bands, as shown in Figs. \ref{Figure 2}(c), \ref{Figure 2}(f), \ref{Figure 2}(i), and \ref{Figure 2}(m). At zero field, the large band gap is clearly revealed in the zero-field DOS in \ref{Figure 2} (c). The low-frequency DOS presents prominent asymmetric peaks, respectively, corresponding to the extremal band-edge states and the partially flat bands along the specific direction in the energy-wave-vector space.
In the presence of an electric field, the splitting of energy bands gives rise to more shoulder-like and peak structures, as illustrated in Figs. \ref{Figure 2}(f), \ref{Figure 2}(i), and \ref{Figure 2}(m). Additionally, the $E_z$-induced constant-energy loops and Dirac cones, respectively, create the asymmetric peaks and valley-like structures, in which the latter appear near the Fermi level. The $E_z$-dependent energy gap and three kinds of Van Hove singularities could be directly identified from  experimental measurements using scanning tunneling spectroscopy (STS)
\cite{GHL;PRL2009,LJY;PRB2016}. Also, STS is the most powerful tool relating the measured tunneling current to the DOS.

\begin{figure}[H]
\centering
{\includegraphics[width=0.9\linewidth]{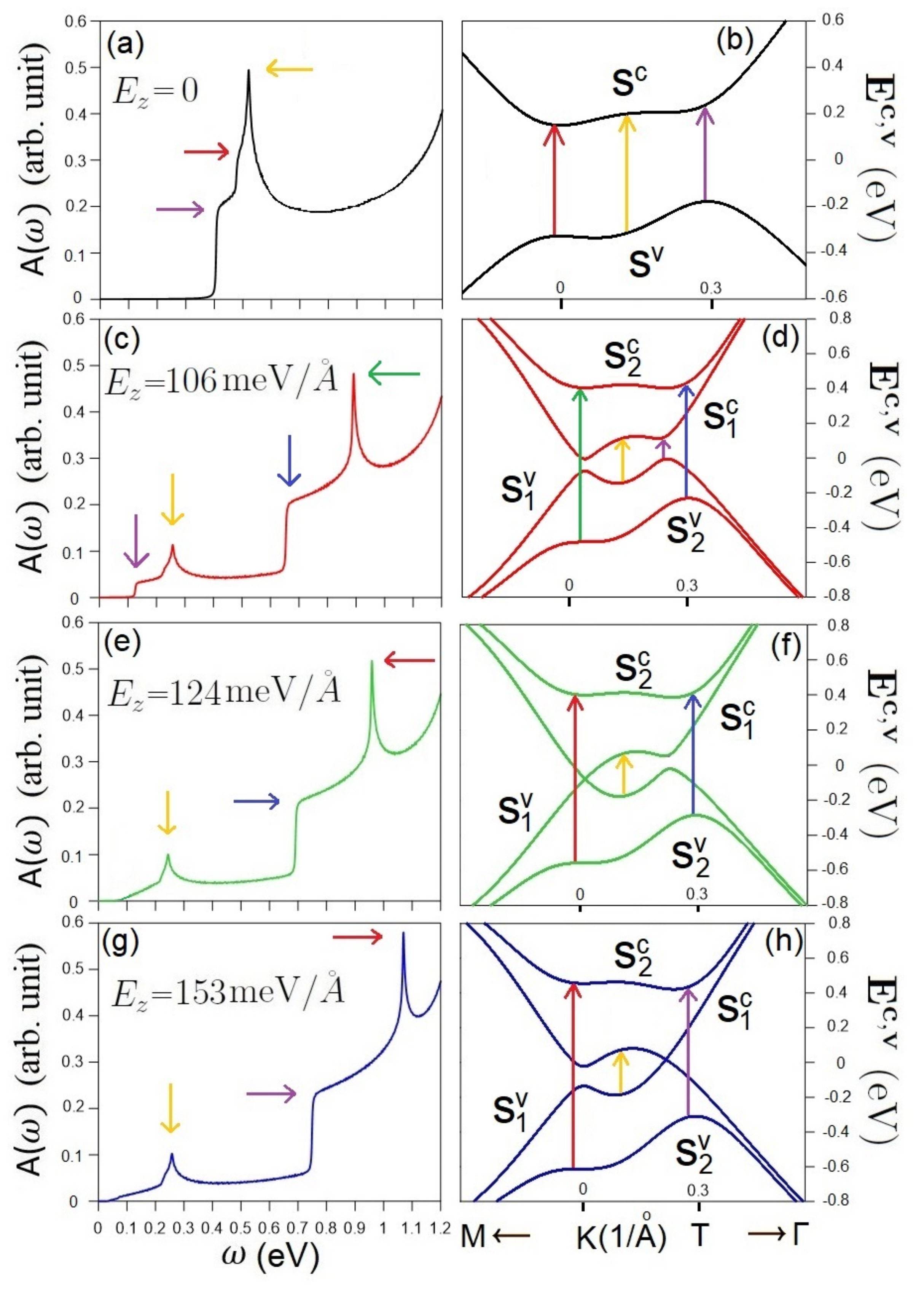}}
\caption{(Color online) The optical-absorption spectra of bilayer silicene in the absence of $E_z$ (a) and the corresponding optical channels (b). Similar figures are shown for (c)-(d) $E_z$ = 106 meV/$\text{\AA}$, (e)-(f) $E_z$ = 124 meV/$\text{\AA}$, and (g)-(h) $E_z$ = 153 meV/$\text{\AA}$.}
\label{Figure 3}
\end{figure}

\medskip
\par

Bilayer silicene exhibits exceptional absorption spectra, in sharp contrast with those of the monolayer system \cite{YB;PRB89}.
There are three special structures in optical excitations, as clearly shown when ${E_z=0}$ by the black curve in Fig. \ref{Figure 3}(a).
The first two belong to the shoulder structures, in which the first and second ones, respectively, arise from the band-edge states at the $\bf{K}$ and $\bf{T}$ points (purple and red arrows in Fig. \ref{Figure 3} and \ref{Figure 3}(b)) . The third structure, the antisymmetric peak (yellow arrow), is due to the weak energy dispersions close to the $\bf{K}$ point. All of them are dominated by Van Hove singularities in the JDOS under vertical excitations. An electric field makes the low-lying optical excitations become more complicated in the presence of the spin-split energy bands. There are two absorption regions, since vertical transitions are allowed only for the same pair of valence and conduction bands (${S_1^v\to\,S_1^c}$ and ${S_2^v\to\,S_2^c}$).
The lower- and higher-frequency absorption regions have similar structures, i.e., the shoulder and peak, when the electric filed is not too strong (e.g., red curve in Fig. 3(c) for ${E_z \leq 106}$ meV/$\text{\AA}$).
The former and the  latter, respectively, come from the optical transitions of the band-edge states near the $\bf{T}$ point and the electronic states from the flat bands near the $\bf{K}$ (Fig. 3(d)). The lowest-frequency threshold shoulder disappears for the higher electric field, e.g, ${E_z=124}$ meV/$\text{\AA}$ and ${E_z=153}$ meV/$\text{\AA}$, because the band-edge states of the first valence/conduction band are also unoccupied/occupied near the $\bf{T}$/$\bf{K}$ point (Figs. 3(e)-3(h)).
The above electric-field-enriched optical structures could be examined by the infrared reflection spectroscopy and absorption spectroscopy \cite{ZQ;NP2008, LM;PRB2008}.

\begin{figure}[H]
\centering
{\includegraphics[width=1.0\linewidth]{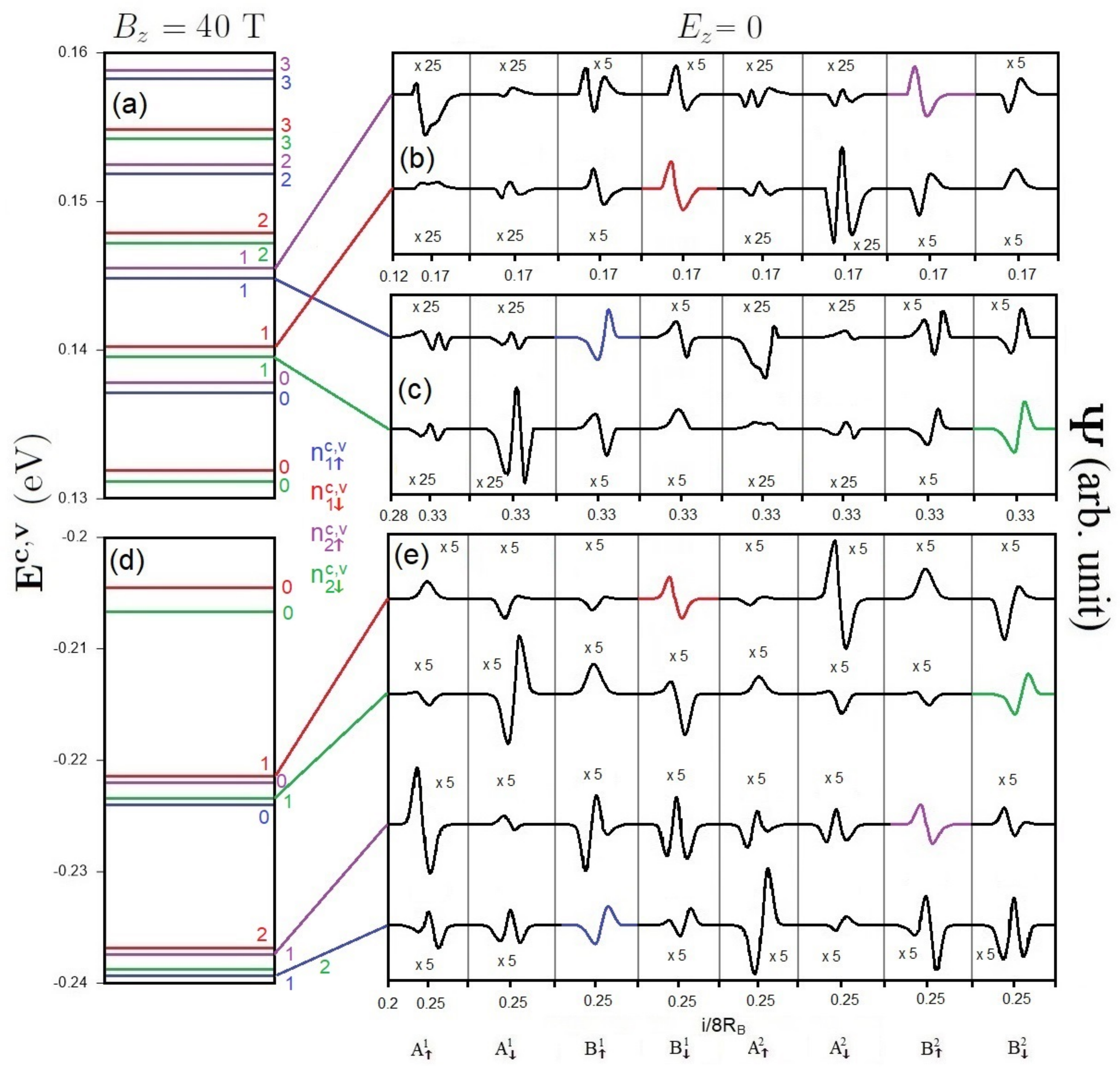}}
\caption{(Color online)  The (a) LL spectrum and (b)-(c) conduction LL wavefunctions on eight distinct sublattices for $B_z$ = 40 T in the absence of electric field. Similar plots for the valence LL spectrum (d) and (e) wavefunctions are also presented.}
\label{Figure 4}
\end{figure}

\begin{figure}[H]
\centering
{\includegraphics[width=1.0\linewidth]{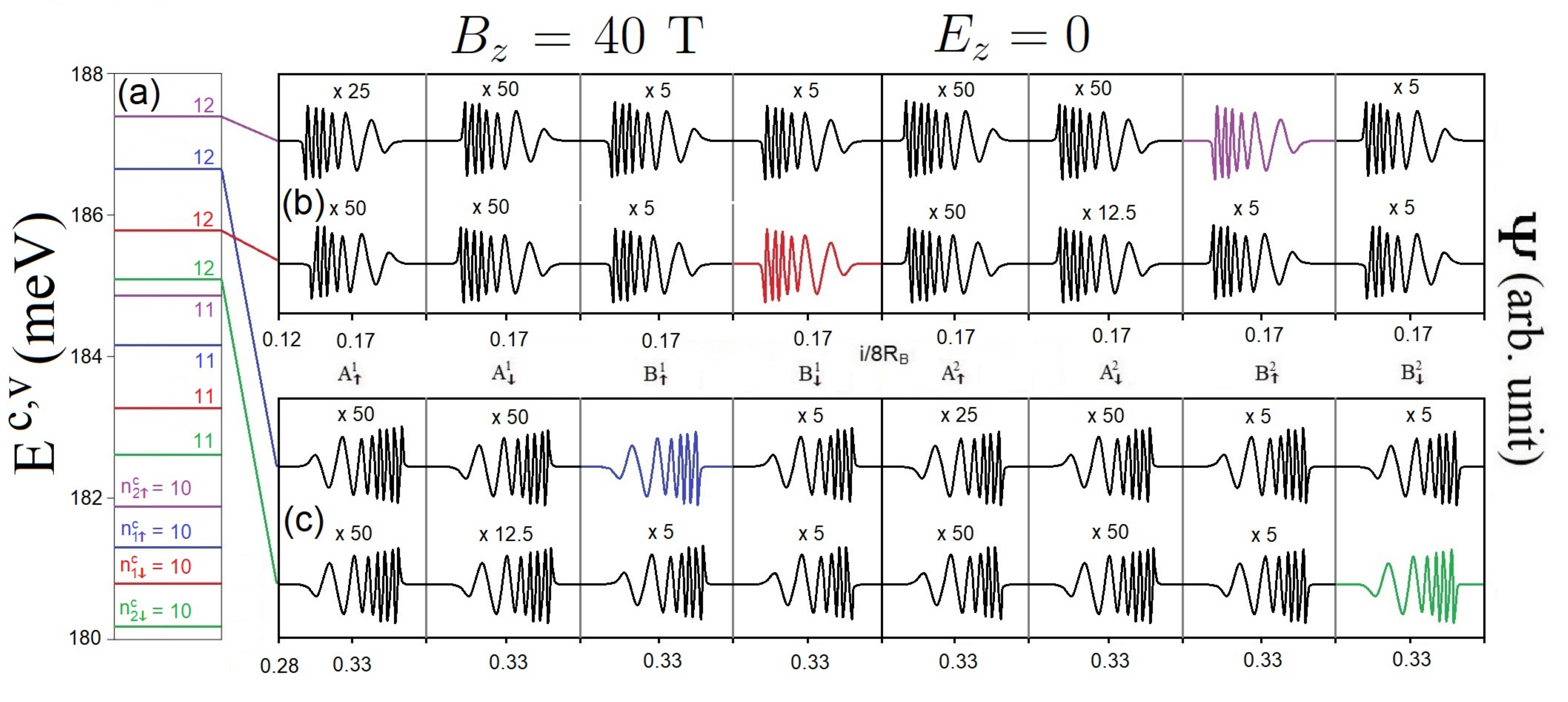}}
\caption{(Color online)  The (a) higher conduction LL spectrum and (b)-(c) LL wavefunctions on eight distinct sublattices for $B_z$ = 40 T in the absence of electric field.}
\label{Figure 5}
\end{figure}

\medskip
\par

AB-bt bilayer silicene exhibits the rich magneto-electronic properties, being thoroughly different from bilayer graphene.
The buckled structure, complex inter-layer atomic interactions, and significant SOCs remarkably enrich the main features of LLs.
The low-lying conduction and valence LLs are quantized from the electronic states near the $\bf{K}$ and $\bf{T}$ points, respectively.
They are doubly degenerate under the interplay of nonequivalent sublattices and SOCs, while there exist the eight-fold degeneracy in bilayer graphene.
The conduction LL wavefunctions are centered at 1/6 (4/6) and 2/6 (5/6) of the expanded unit cell while the valence ones are localized at 1/4 (3/4) center \cite{TND;Si}.
Such wavefunctions are the well-behaved spatial distributions characterized by the ${3p_z-}$ and spin-dependent sub-envelope functions on the eight sublattices, as demonstrated in Fig. 4.
Their quantum numbers are defined by the number of zero points of the spatial probability distributions in the dominated sublattices.
Furthermore, they are very useful in understanding the magneto-optical selection rules of the inter-LL transitions.
In principle, LLs can be classified as four distinct subgroups ($n^{c}_{\uparrow 1}$, $n^{c}_{\downarrow 1}$, $n^{c}_{\uparrow 2}$ and $n^{c}_{\downarrow 2}$) based on the sublattice- and spin-dominated wavefunctions (blue, red, purple, and green lines in Figs. 4(a) and 4(d)).
Four LL subgroups possess the usual orderings of state energy and energy spacing; that is, such properties, respectively, grow and decline with the increase/decrease of $E^c$/$E^v$.
The LL energy splitting is induced by both SOCs and stacking configuration/interlayer atomic interactions, in which the former is much larger than the later.
The split energy is strongly dependent on the magnetic field strength (discussed later in Fig. 7(a)).
\medskip
\par

Since the low-lying valence and conduction have different localization centers, the vertical magneto-optical transitions between them are forbidden.
The higher conduction and deeper valence LLs, with many oscillation modes, are critical in understanding the magneto-optical excitations; therefore, they deserve a closer observation.
They are, respectively, quantized from the electronic states near $\bf{T}$ and $\bf{K}$ valleys, corresponding to the shoulder-like energy bands (Fig. 2(b)). This is thoroughly different from the magnetic quantization in bilayer graphene systems \cite{YKH;SR2014}.
The energy spectrum and spatial distributions of the higher conduction LLs are clearly illustrated in Figs. 5(a)-(c))
They clearly exhibit the non-symmetric and non-well-behaved spatial distributions; that is, they belong to the perturbed LLs with the main and the side modes \cite{YKH;SR2014}.
For the higher conduction/deeper valence LLs, their contribution widths are wide and the effective width covers two localization centers of (1/6, 1/4) $\&$ (2/6, 1/4). As the result, the low-lying conduction/valence and the deeper valence/higher conduction LLs will have a significant overlap in the spatial distributions, being expected to be very important in the magneto-optical threshold excitation.

\begin{figure}[H]
\centering
{\includegraphics[width=1.0\linewidth]{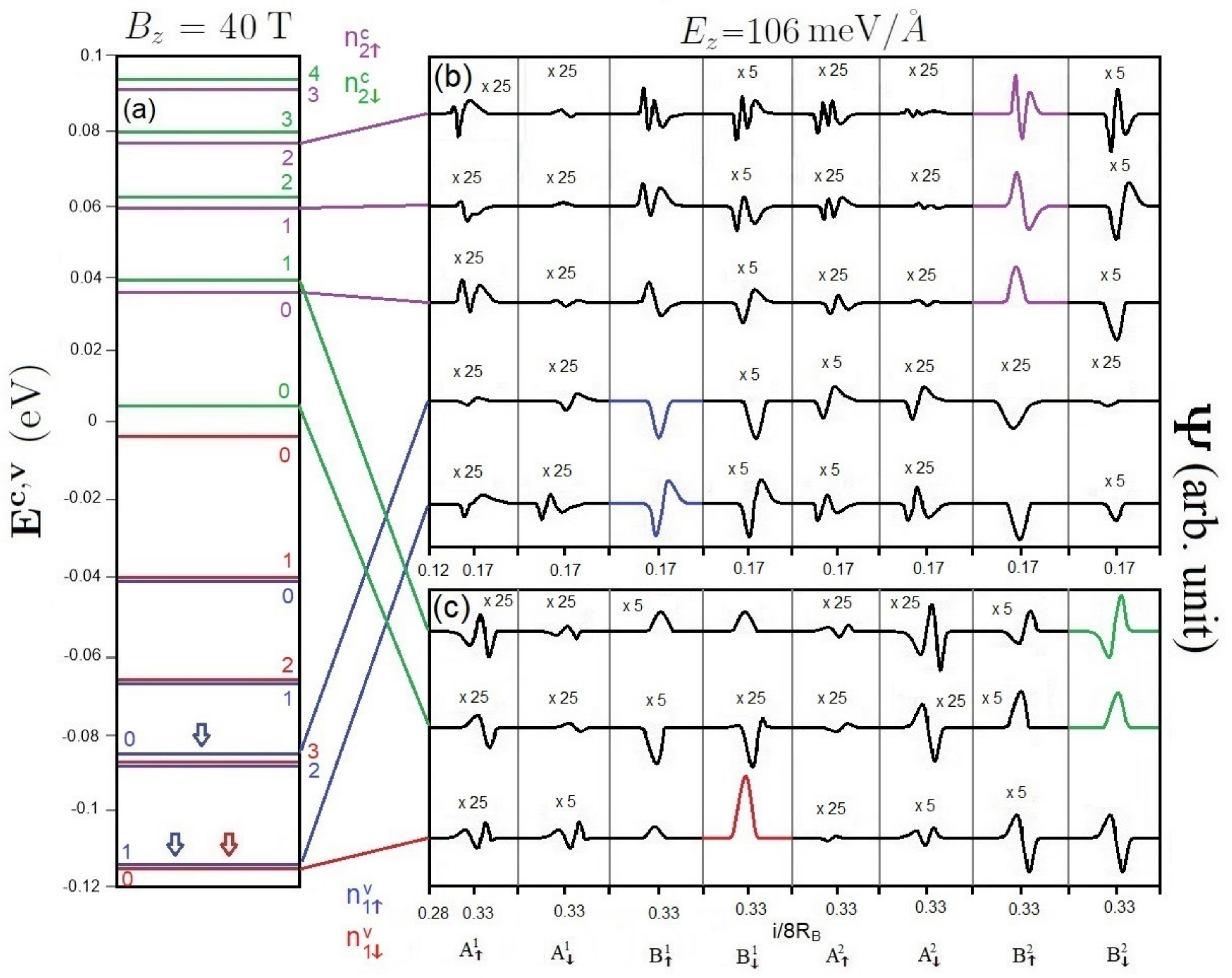}}
\caption{(Color online)  The (a) LL spectrum and (b)-(d) LL wavefunctions on eight distinct sublattices for $B_z$ = 40 T under $E_z$ = 106 meV/$\text{\AA}$.}
\label{Figure 6}
\end{figure}

\begin{figure}[H]
\centering
{\includegraphics[width=1.0\linewidth]{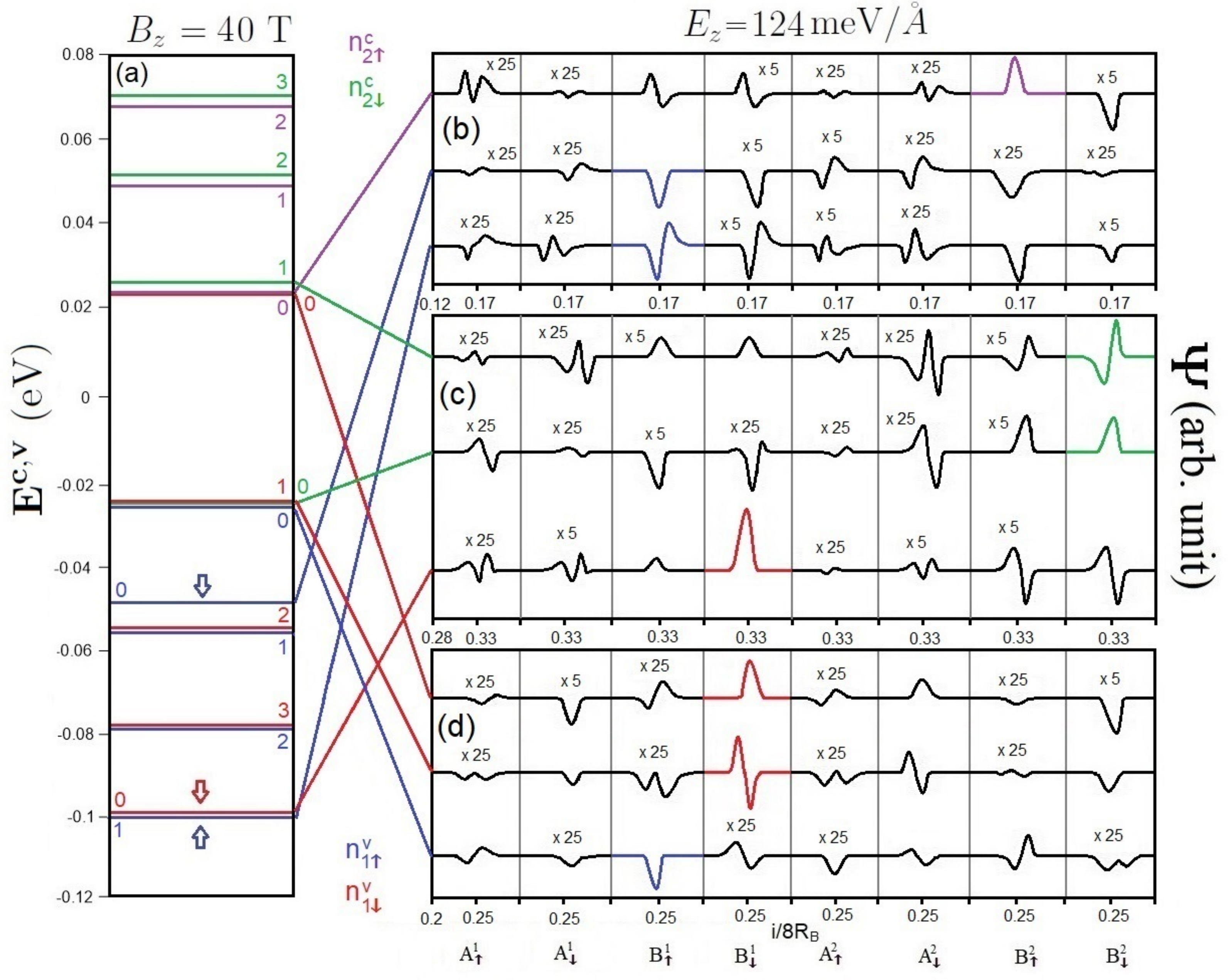}}
\caption{(Color online)  The (a) LL spectrum and (b)-(d) LL wavefunctions on eight distinct sublattices for $B_z$ = 40 T under $E_z$ = 124 meV/$\text{\AA}$.}
\label{Figure 7}
\end{figure}
\medskip
\par

The magneto-electronic properties can be greatly diversified by an external electric field.
For the zero-gap band structure under a critical electric field, $E_z$ = 106 meV/$\text{\AA}$, the magnetic quantization is initiated from the valence and conduction states near the $\bf{K}$ and $\bf{T}$ valleys (Fig. 6).
There exist the low-lying valence and conduction LLs with the same localization center simultaneously, instead of only conduction or valence ones as for $E_z$ = 0.
In particular, the well-behaved valence LLs at 1/6 and 2/6 centers (the conduction ones at 1/4 center) come to exist, as marked by the arrows in Fig. \ref{Figure 6}(a).
Such LLs come from the rather pronounced oscillating band structure in the presence of an electric field (Fig. 2(e)).
The LL energy spectrum and spatial distributions for this critical electric field are clearly illustrated in Figs. \ref{Figure 6}(a)-\ref{Figure 6}(c).
Four subgroups of LLs do not appear together, but are separated into two lower- and higher/deeper-energy ones. This means that, the splitting related to the non-equivalence of sublattices is greatly enhanced by the electric field because of the distinct Coulomb site energies.
We only focus on the former associated with the lower-frequency magneto-optical excitations. For each valley, the two subgroups of  low-lying conduction and valence LL are, respectively, corresponding to ($n^{c}_{\uparrow 2}$, $n^{c}_{\downarrow 2}$) and ($n^{c}_{\uparrow 1}$, $n^{c}_{\downarrow 1}$).
It should be noticed that, they are dominated by the different sublattices.
The vertical transitions, the valence to conduction LLs, from the different valleys have much contribution to the magneto-optical spectra.

\medskip
\par

On the other hand, a perpendicular electric field may lead to the formation of Dirac cones at the $\bf{K}$ or $\bf{T}$ valleys, giving rise to the special LL quantization.
For the second critical electric field, $E_z$ = 124 meV/$\text{\AA}$, the conduction and valence LLs which are initiated from the $\bf{K}$ and $\bf{T}$ valleys, respectively, correspond to ($n^{c}_{\downarrow 2}$, $n^{v}_{\uparrow 2}$) and ($n^{c}_{\downarrow 1}$, $n^{v}_{\uparrow 1}$), in which all the LL wavefunctions are well-behaved in the spatial distribution, as clearly shown in Fig. 7. The former is similar to those from the linear Dirac cone \cite{CY;IOP2017}.
Their energies are characterized by $E^{c (v)}_{2 (1)}$ $\propto \sqrt{n^{c}_{\downarrow 2} (n^{v}_{\uparrow 1})}$, a simple relation is absent in the latter.
Specifically, the energy spacing of $n^{c}_{\downarrow 2}$ = 0 and $n^{v}_{\uparrow 1}$ = 0 is finite and gradually grows with the magnetic field strength, such as the magnitude of 25 meV at $B_z$ = 40 T.
The opposite is true for the third critical electric field, $E_z$ = 153 meV/$\text{\AA}$.
That is, the Dirac cone-like LLs are created by the electronic states near the $\bf{T}$ valley.
The sharp contrast between the $\bf{K}$ and $\bf{T}$ valleys will be directly reflected in the magneto-optical excitations.
Moreover, there exist certain important differences compared to monolayer graphene with a zero energy spacing between the $n^c$ = 0 and $n^v$ = 0 LLs, eight-fold degenerate LLs, and the same dominating sublattices for the valence and conduction LLs \cite{CY;IOP2017}.

\medskip
\par

\begin{figure}[H]
\centering
{\includegraphics[width=1.0\linewidth]{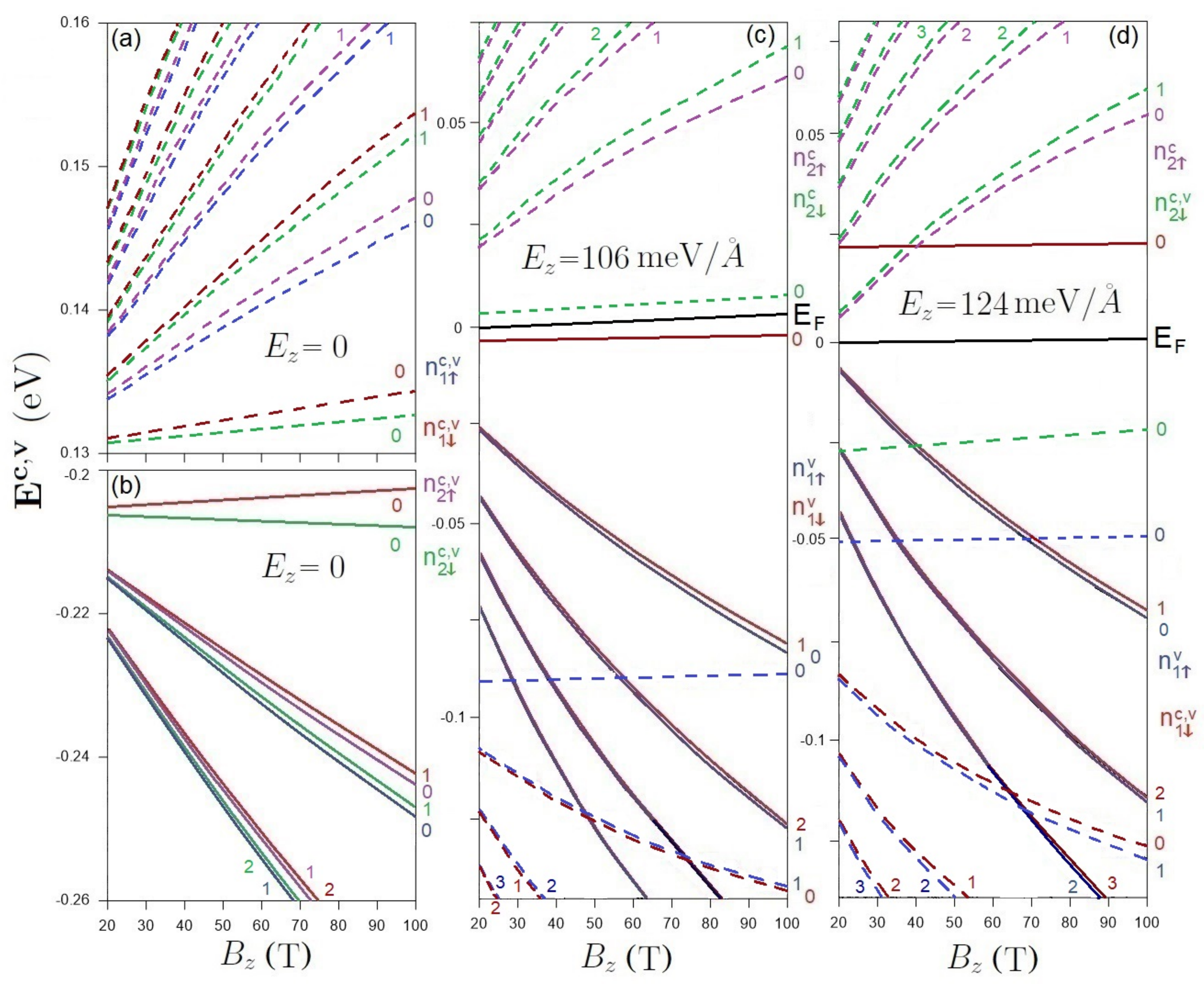}}
\caption{(Color online)  The $B_z$-dependent LL energy spectra in the absence of electric field (a) and under (b) $E_z$ = 106 meV/$\text{\AA}$ and (c) $E_z$ = 124 meV/$\text{\AA}$. The LLs near the $\bf{K}$ and $\bf{T}$ valleys are presented by the dashed and solid curves, respectively.}
\label{Figure 8}
\end{figure}
\medskip
\par

The $B_z$-dependent LL energy spectrum is very useful for the comprehension of the magnetic quantization.
Without an electric field, the conduction/valence LL energies are generally increased/declined with the growth of magnetic field strength, as clearly shown in Fig. 8(a) by the dashed and solid curves (initiated from the $\bf{K}$ and $\bf{T}$ valleys). Exceptionally, the initial valence $n^{v}_{\downarrow 1}$ = 0 LL energy slowly grows with $B_z$. The energy gap, which is determined by the $n^{c}_{\downarrow 2}$ = 0 and $n^{v}_{\downarrow 1}$ = 0 LLs, almost remains the same ($E_g (B_z) \approx$ 340 meV).
The valence and conduction LL energy spectrum is asymmetric about the Fermi level, mainly owing to the important interlayer hopping integrals and layer-dependent SOCs.
Four subgroups of LLs behave similarly during the variation of $B_z$, in which the low-lying spectrum presents the almost linear $B_z$-dependence, except for the initial valence $n^{v}_{\downarrow 1}$ = 0 LL near the $\bf{T}$ valley.
This reflects the low-lying parabolic dispersions near the $\bf{K}$ and $\bf{T}$ valleys (Fig. 2(b)).

\medskip
\par

It should be noticed that, the onset energies of LL subgroups are strongly dependent on the strength of electric field.
The effects of composite fields are clearly illustrated by the $B_z$-dependent LL energy spectra at the critical electric field strengths, as shown in Figs. \ref{Figure 8}(b) and \ref{Figure 8}(c).
The asymmetry of LL energy spectrum is enhanced under the first critical electric field, $E_z$ = 106 meV/$\text{\AA}$ (Fig. \ref{Figure 8}(b)).
There are four initial LLs of ($n^{v}_{\uparrow 1}$ = 0 $\&$ $n^{c}_{\downarrow 2}$ = 0) from the $\bf{K}$ valley and ($n^{v}_{\downarrow 1}$ = 0 $\&$ $n^{c}_{\uparrow 2}$ = 0) from the $\bf{T}$ valley which present the weak $B_z$-dependence, as demonstrated in Fig. 8(b).
Specifically, the above-mentioned $n^{c}_{\downarrow 2}$ = 0 and $n^{v}_{\downarrow 1}$ = 0 LLs determine the Fermi level, being the middle of the nearest occupied and unoccupied LLs.
The energy gap, the energy spacing between these both LLs becomes very narrow and grows with the magnetic field. It almost vanishes at sufficiently low $B_z$.
According to the numerical examinations, the energy spacing and magnetic field strength do not present a simple linear nor square-root relationships.
In addition, there is only a few well-behaved conduction LLs near the $\bf{T}$ valley ($n^{c}_{\uparrow2}$ $\&$ $n^{c}_{\downarrow2}$) (only $n^{c}_{\uparrow2}$ = 0 LLs is observed for $B_z \geq$ 20 T) and they are located at relatively high energy, compared with those initiated from the $\bf{K}$ valley.
The latter will dominate the threshold magneto-optical excitations.

\medskip
\par

Under the second critical electric field of $E_z$ = 124 meV/$\text{\AA}$, the conduction and valence LL subgroups initiated from the $\bf{K}$ valley approach to each other, as shown in Fig. 8(c).
This directly reflects the linear and isotropic Dirac-cone structures near the $\bf{K}$ point (Fig. 2(h)).
As for the LLs quantized from the $\bf{T}$ valley, only the valence ones come to exist near the Fermi level, in which the $n^v_{\downarrow 1}$ = 0 LL (the red solid curve) is even high than few $n^{c}_{\downarrow2}$ conduction LLs from the $\bf{K}$ valley.
Regarding the former, the zero-quantum-number LLs of $n^{c}_{\downarrow 2}$ and $n^{v}_{\downarrow 1}$ reach the minimum energy spacing.
Furthermore, the LL energies and the magnetic field strength possess a specific relationship of $E^{c (v)}_{2 (1)}$ $\propto \sqrt{B_z}$, similarly to that in monolayer graphene \cite{CY;IOP2017}.
Generally, the initial LLs of ($n^{v}_{\uparrow 1}$ = 0 $\&$ $n^{c}_{\downarrow 2}$ = 0) and ($n^{v}_{\downarrow 1}$ = 0, respectively, from the $\bf{K}$ and $\bf{T}$ valleys are hardly affected by the magnetic field strength.
The Fermi level is determined by the ($n^{c}_{\downarrow 2}$ = 0 $\&$ $n^{v}_{\downarrow 1}$ = 0) LLs at sufficiently high magnetic field strength ($B_z \leq$ 40 T) and ($n^{c}_{\uparrow 2}$ = 0 $\&$ $n^{v}_{\downarrow 1}$ = 1) ones for smaller $B_z$.
However, the value of $E_F$ is almost independent of $B_z$.
Especially, an electric field can alter certain LLs nearest to the Fermi level, leading to dramatic changes in the magneto-optical threshold channels.
For example, the valence $n^{v}_{\downarrow 1}$ = 0 LL from the $\bf{T}$ valley and conduction $n^{c}_{\uparrow 2}$= 0 LL from the $\bf{K}$ one becomes unoccupied and occupied, respectively, as illustrated in Fig. \ref{Figure 8}(c).
The above-mentioned features are also revealed in the $B_z$-dependent LL energy spectra for the third critical field of $E_z$ = 153 meV/$\text{\AA}$ by means of the interchange of the $\bf{K}$ and $\bf{T}$ valleys.
In general, an electric field creates more low-lying well-behaved conduction and valence LLs and thus is expected to induce very complicated magneto-optical absorption spectra.

\begin{figure}[H]
\centering
{\includegraphics[width=1.0\linewidth]{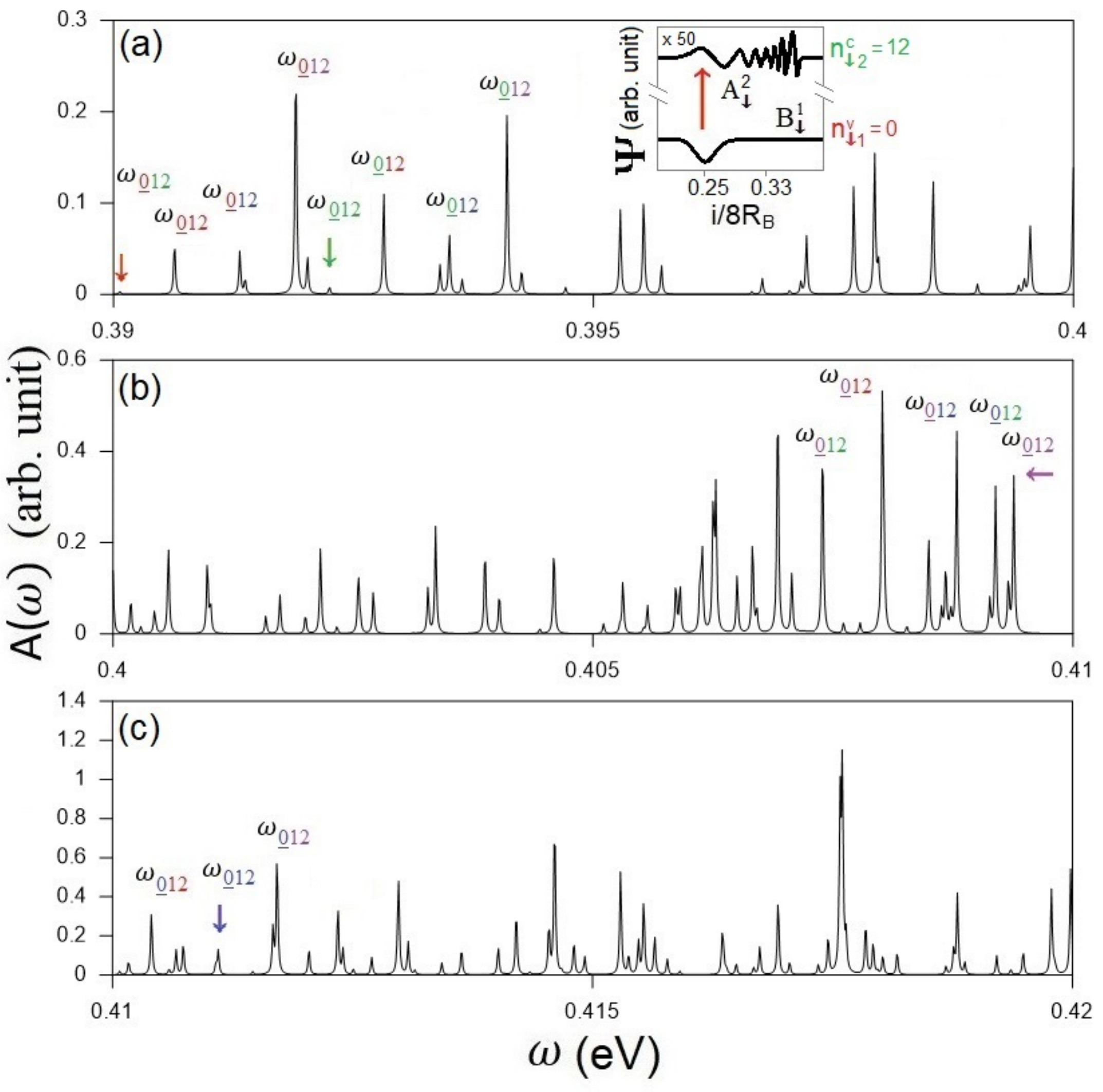}}
\caption{(Color online)  The spectral intensities of AB-bt bilayer silicene for $B_z$ = 40 T in the absence of electric field.}
\label{Figure 9}
\end{figure}
\medskip
\par

AB-bt bilayer silicene presents the feature-rich magneto-absorption spectra, reflecting the unusual LLs and band structures.
The vertical transitions among four subgroups of LLs lead to many single, double and twin delta-function-like peaks with non-uniform intensities, as shown in Figs. 9(a)-9(c).
There are 4 $\times$ 4 categories of inter-LL optical transitions, covering 4 intra-subgroup (part of peaks are marked by the arrows with distinct colors) and 12 inter-subgroup ones.
According to the magneto-optical absorption functions \cite{CY;IOP2017}, the inter-LL transition is available whenever the initial and final states related to the two sublattices in the large $(t_0, t_1)$ hopping integrals possess the same quantization mode.
As a result of an indirect energy gap, in each category, absorption peaks are corresponding to the optical transitions associated with the multi-mode LLs in the absence of a specific selection rule.
This results in a lot of magneto-absorption peaks even within a very narrow-range frequency of ${\sim\,30}$ meV, being never observed in the other condensed-matter systems \cite{CY;IOP2017, ZQ;NP2008, LM;PRB2008, MKF;PRL2010, LZQ;PRL2009, JZ;PRL2007, PP;PRL2008, MKF;PRL2009, MKF;PRL2008, PP;PRB2012, NRJ;PRL2013, OM;PRB2011}.
For a LL with sufficiently large quantum number, the extended oscillation wavefunctions localized at (1/4)/(1/6 and 2/6) along the x-axis will overlap with that of another LL at the neighboring localization centers of (1/6 and 2/6)/(1/4). For example, the spatial distributions of the $n_{1\downarrow}^v = 0$ and $n_{1\downarrow}^c = 12$ LLs are illustrated in the inset of Fig. 9(a). The former and the latter are localized at 1/4 and 1/6 centers which are very close to each other, leading to an obvious overlapping phenomenon between them.
This enables the vertical optical transitions between the initial- and final-state LLs near the $\bf{T}$ ($\bf{K}$) valley, in which the magneto-absorption peaks have the different intensities and frequencies, e.g., the 16 absorption structures due to the ($n_{1\downarrow}^v = 0$, $n_{1\uparrow}^v = 0$, $n_{2\uparrow}^v = 0$, $n_{2\downarrow}^v = 0$) and ($n_{1\downarrow}^c = 12$, $n_{1\uparrow}^c = 12$, $n_{2\uparrow}^c = 12$, $n_{2\downarrow}^c = 12$) LLs.
The very special magneto-optical property is absent in other well-known 2D systems, e.g., graphene \cite{CY;IOP2017} and phosphorene \cite{JYW;X2018}.
The threshold absorption peak, the optical gap, belongs to the intra-subgroup $n_{1\downarrow}^v = 0$ $\rightarrow$ $n_{1\downarrow}^c = 12$ transition (red arrow in Fig. 9(a)).
Its frequency is expected to be dependent on both magnetic and electric fields.
Especially, the optical gap ($\approx$ 0.39 eV for $B_z$ = 40 T) is larger than the energy gap ($\approx$ 0.3 eV ), since the vertical transitions between the low-lying conduction and valence LLs at different centers are forbidden.
In addition, there also exist the 16 excitation categories related to the initial conduction LLs from the $\bf{K}$ valley, such as, those associated with the ($n_{1\downarrow}^v = 12$, $n_{1\uparrow}^v = 12$, $n_{2\uparrow}^v = 12$, $n_{2\downarrow}^v = 12$) and ($n_{1\downarrow}^c = 0$, $n_{1\uparrow}^c = 0$, $n_{2\uparrow}^c = 0$, $n_{2\downarrow}^c = 0$) LLs. However, they do not contribute to the threshold magneto-optical excitation.

\begin{figure}[H]
\centering
{\includegraphics[width=1.0\linewidth]{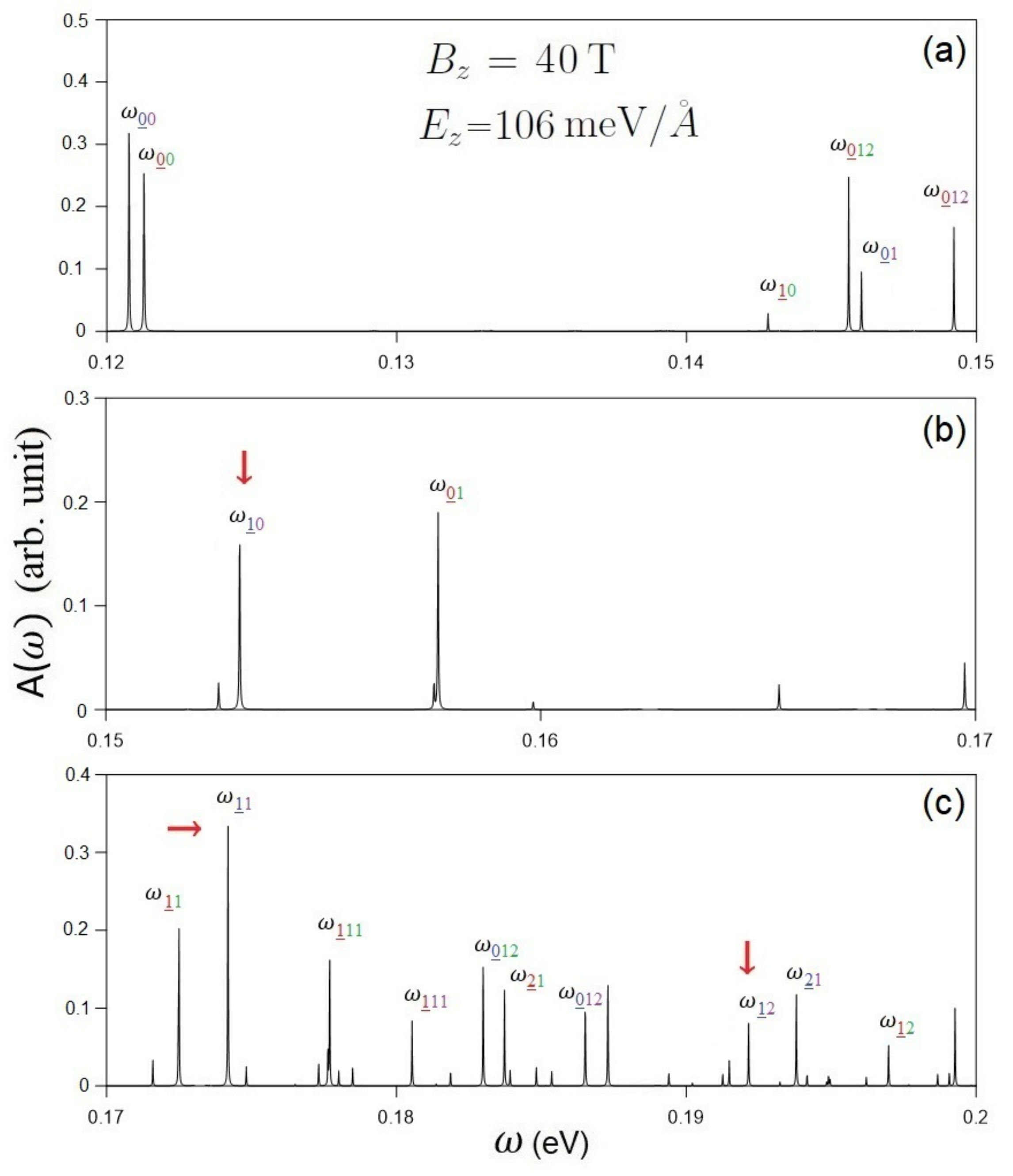}}
\caption{(Color online)  The spectral intensities of AB-bt bilayer silicene for $B_z$ = 40 T under $E_z$ = 106 meV/$\text{\AA}$.}
\label{Figure 10}
\end{figure}
\medskip
\par

The magneto-optical spectra are diversified under the interplay of electric field, magnetic field, significant SOCs and interlayer atomic interactions.
An applied electric field can create the inter-LL optical transitions at lower frequency, and they satisfy the specific selection rules.
There exist the optical excitations related to LLs from the same localization center (type I) and the neighboring ones (1/4 and 1/6 (2/6) centers) (type II).
The type-I absorption peaks are associated with the $E_z$-induced new low-lying well-behaved valence LLs at the $\bf{K}$ valley.
They are only characterized by the LLs with the same spin configuration ($n^{v}_{\uparrow 1}$ $\rightarrow$ $n^{c}_{\uparrow 2}$ and $n^{v}_{\downarrow 1}$ $\rightarrow$ $n^{c}_{\downarrow 2}$).
The threshold peak, which is determined by the $n^{v}_{\uparrow 1}$ = 0 $\rightarrow$ $n^{c}_{\uparrow 2}$ = 0 transition, is located at much lower frequency ($\omega_{th} \approx$ 120 meV) compared with that at zero electric field.
It should be noticed that, the type-I magneto-optical excitations obey the optical selection rules of $\Delta n$ = 0 and $\pm$ 1, as demonstrated in Fig. 10 for the first critical electric field ($E_z$ = 106 meV/$\text{\AA}$) under $B_z$ = 40 T.
This is because each LL possesses a main mode and a few side modes, referring to Fig. 6.
For example, the $n^{v}_{\uparrow 1}$ = 1 LL contains a main mode of 1 on the dominating $B^1_{\uparrow}$ sublattice and the side modes of 0 and 2 on the other sublattices (Fig. 6(b)).
As a result, there are the available inter-LL optical transitions of $n^{v}_{\uparrow 1}$ = 1 $\rightarrow$ $n^{c}_{\uparrow 2}$ = 0, $n^{v}_{\uparrow 1}$ = 1 $\rightarrow$ $n^{c}_{\uparrow 2}$ = 1, and $n^{v}_{\uparrow 1}$ = 1 $\rightarrow$ $n^{c}_{\uparrow 2}$ = 2, as marked by the red arrows in Figs. 10(b) and 10(c).
As for the type-II absorption peaks, there are excitation channels between n = 0 and 12 LLs, similarly to those in the absence of electric field; that is, they are associated with the low-lying well-behave valence LLs from the $\bf{T}$ valley.
The above-mentioned absorption peaks, including the type-I and type-II ones, belong to the 4 excitation categories, but not 16 ones. They cover the $n^{v}_{\uparrow 1}$ $\rightarrow$ $n^{c}_{\uparrow 2}$, $n^{v}_{\uparrow 1}$ $\rightarrow$ $n^{c}_{\downarrow 2}$, $n^{v}_{\downarrow 1}$ $\rightarrow$ $n^{c}_{\uparrow 2}$, and $n^{v}_{\downarrow 1}$ $\rightarrow$ $n^{c}_{\downarrow 2}$ magneto-optical transitions.

\begin{figure}[H]
\centering
{\includegraphics[width=1.0\linewidth]{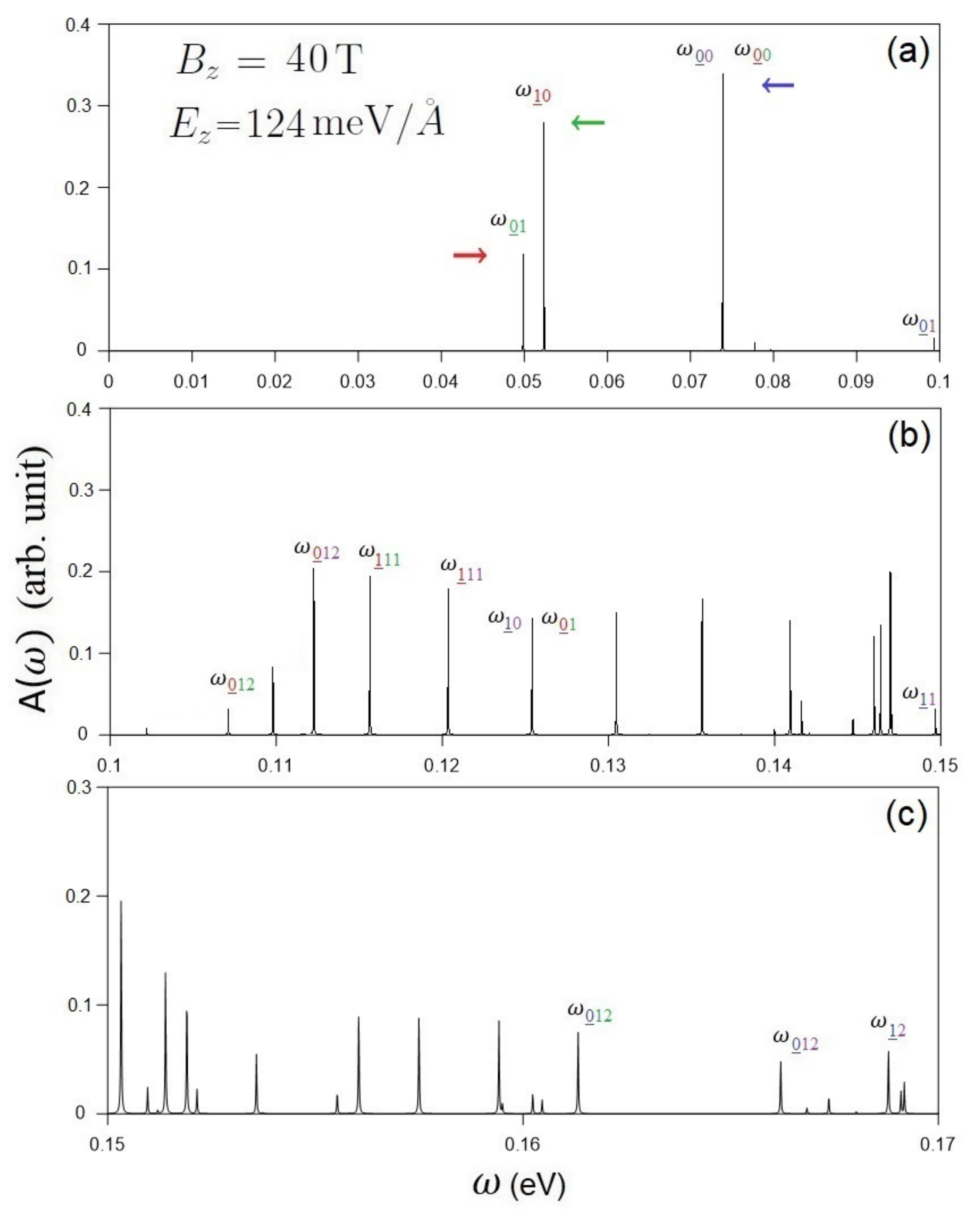}}
\caption{(Color online)  The spectral intensities of AB-bt bilayer silicene for $B_z$ = 40 T under $E_z$ = 124 meV/$\text{\AA}$.}
\label{Figure 11}
\end{figure}
\medskip
\par

It is worth to consider the magneto-optical spectra at the second critical electric field ($E_z$ = 124 meV/$\text{\AA}$) when a Dirac cone is formed at the $\bf{K}$ valley.
Both type-I and type-II magneto-optical excitations come to exist in the absorption spectra.
The former satisfy the optical selection rules of $\Delta n$ = 0 and $\pm$ 1 while the later do not.
Especially, the formation of Dirac cone induces the extraordinary phenomena.
Since the Dirac cone lies below the Fermi level (Fig. 2(h)), the occupation of some LLs near the Dirac point is altered, which directly affects the threshold magneto-optical excitation (Fig. 11).
The intra-subgroup excitation channels of $n^{c}_{\downarrow 2}$ = 0 $\rightarrow$ $n^{c}_{\downarrow 2}$ = 1 (red arrow) and $n^{v}_{\downarrow 1}$ = 1 $\rightarrow$ $n^{v}_{\downarrow 1}$ = 0 (green arrow) comes into existence because the conduction $n^{c}_{\downarrow 2}$ = 0 LL near the $\bf{K}$ valley and $n^{v}_{\downarrow 1}$ = 0 one near the $\bf{T}$ valley become occupied and unoccupied, respectively.
The threshold absorption peak, which is determined by the former, is present at rather low frequency ($\omega_{th} \sim$ 50 meV).
Furthermore, there exists a double peak due to the two peaks of ($n^{v}_{\uparrow 1}$ = 0 $\rightarrow$ $n^{c}_{\uparrow 2}$ = 0 and $n^{v}_{\downarrow 1}$ = 0 $\rightarrow$ $n^{c}_{\downarrow 2}$ = 0) merging together in the absorption spectrum, as indicated by the blue arrow in Fig. 11(a). The crossing behavior between LLs is responsible for this double peak.
In general, the type-I magneto-optical excitations are mainly revealed near the $\bf{K}$ valley, except for the $n^{v}_{\downarrow 1}$ = 1 $\rightarrow$ $n^{v}_{\downarrow 1}$ = 0 (green arrow) near the $\bf{T}$ valley.
Similar magneto-optical properties can also be observed for the third critical electric field ($E_z$ = 153 meV/$\text{\AA}$) when the Dirac cone at the $\bf{T}$ valley is located above the Fermi level.

\begin{figure}[H]
\centering
{\includegraphics[width=0.9\linewidth]{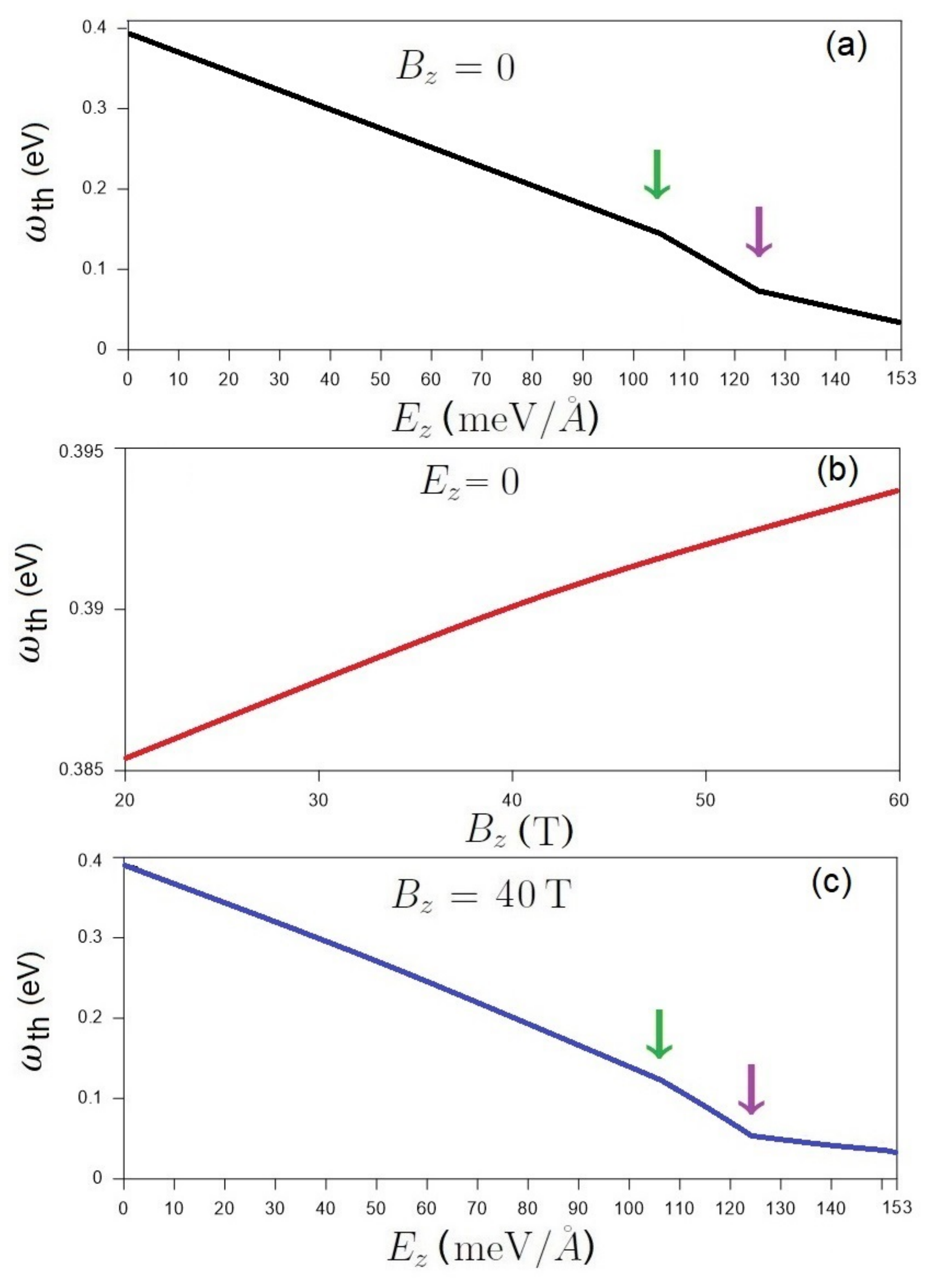}}
\caption{(Color online)  The (a) $B_z$-dependent threshold frequencies at zero electric field and (b) $E_z$-dependent threshold frequencies for $B_z$ = 40 T.}
\label{Figure 12}
\end{figure}
\medskip
\par

The optical gap strongly depends on the strength of electric and magnetic fields, as demonstrated in Figs. 12(a)-12(c).
With the increase of electric field up to the first critical field ($E_z$ = 106 meV/$\text{\AA}$), the threshold frequency, which is determined by the first shoulder-like absorption structure (Fig. 3(a)), is gradually declined from around 0.4 eV, as clearly illustrated in Fig. 12(a).
Right after this field, such structure is absent since the conduction/valence band-edge state near the Fermi level becomes occupied/unoccupied for the $\bf{K}$/$\bf{T}$ valley. As a result, the threshold frequency, being associated with the excitation of electronic states near the $\bf{K}$ valley, continues to decrease. The further increase of $E_z$ changes the threshold excitation to be near the $\bf{T}$ valley and slightly lowers the optical gap.
Regarding the magneto-optical threshold peak in the absence of electric field, its frequency is characterized by the type-II absorption peak of $n_{1\downarrow}^v = 0$ $\rightarrow$ $n_{1\downarrow}^c = 12$ (red arrow in Fig. 9(a)).
The optical gap monotonically grows with the increment of magnetic field, as clearly shown in Fig. 12(b).
This is in a consistency with the $B_z$ evolution of LL energies, in which the conduction/valence LL energies gradually rise/decline with increasing $B_z$ (Fig. 8(a)).
Under the composite fields, the threshold frequency has an inverse relation with the electric field for a fixed magnetic field, as demonstrated in Fig. 12(c) for $B_z$ = 40 T.
With increasing $E_z$ from zero up to 153 meV/$\text{\AA}$, the optical gap is generally decreased.
For $E_z <$ 106 meV/$\text{\AA}$ where the band gap is nonzero, the threshold peak is determined by the type-II $n_{1\downarrow}^v = 0$ $\rightarrow$ $n_{1\downarrow}^c = 12$ excitation channel; the optical gap is monotonically declined with $E_z$. After that, the threshold peak belongs to the type-I excitation channel of $n^{c}_{\downarrow 2}$ = 0 $\rightarrow$ $n^{c}_{\downarrow 2}$ = 1 at the $\bf{K}$ valley (red arrow in Fig. 11(a)); the optical gap declines more quickly (from the green arrow to the purple one). As for greater $E_z$, the threshold peak is corresponding to the type-I excitation channels at the $\bf{T}$ point and its frequency decreases slowly.

\medskip
\par

There are certain important differences between bilayer AB-bt silicene and AB-stacked graphene in magneto-electronic and optical properties \cite{CY;IOP2017}. For the latter, all the LLs possess eight-fold degeneracy without sublattice non-equivalence and spin splitting, mainly owing to the absence of the buckled structure, very significant interlayer hopping integrals, and important SOCs. Their localization centers are only present at 1/6 (4/6) and 2/6 (5/6), but absent at 1/4 (3/4). They are not affected by an electric field. However, the state degeneracy is reduced to half in the presence of $E_z$, since the inversion symmetry/bi-sublattice equivalence is destroyed by the Coulomb potential site energies. The sublattice-dependent LL energy spectra  exhibit the diverse behaviors, anti-crossing, crossing $\&$ non-crossing behaviors during the variation of $B_z$/$E_z$. However, there are no low-lying anti-crossing spectra in silicene systems. Concerning the first group of valence and conduction LLs, the available magneto-excitation category/categories is/are one/two in the absence/presence of $E_z$. Obviously, the magneto-absorption peaks are well characterized by the specific selection rule of $\Delta n$ = $\pm$ 1 except that the extra $\Delta n$ = 0 $\&$ 2 rules might appear under a perpendicular electric field. Their quantum numbers are much smaller than those in bilayer silicene. The threshold channel is only associated with the small quantum-number LLs; that is, it is determined by ${n^v=0/1/2}$ and ${n^c=1/0/1}$

\medskip
\par

Absorption \cite{MKF;PRL2010, LZQ;PRL2009}, transmission \cite{LZQ;PRL2009, JZ;PRL2007, PP;PRL2008, MKF;PRL2009} and reflection spectroscopies \cite{LZQ;PRL2009, MKF;PRL2008} are the most efficient techniques in exploring the essential optical excitations of condensed-matter systems. They are employed as analytical tools for a characterization of optical properties, when the experimental measurements are taken on the fraction related to the adsorbed, transmitted, or reflected light by a sample within a desired frequency range. A broadband light source is utilized and done through a tungsten halogen lamp with the broad range of modulation intensity and frequency \cite{JZ;PRL2007, MKF;PRL2008}. The transmission experiments have confirmed that the absorption intensity of monolayer graphene is proportional to the frequency because of the linear dispersions in the isotropic Dirac cone of massless fermions \cite{JZ;PRL2007}. Besides, massive Dirac fermions are identified in AB-stacked bilayer graphene \cite{LZQ;PRL2009, MKF;PRL2009}. Moreover,  infrared reflection and absorption spectroscopies are also utilized to verify the partially flat and sombrero-shaped energy bands of ABC-stacked few-layer graphenes \cite{MKF;PRL2010}. Three kinds of optical spectroscopies are very useful to examine the stacking- and $E_z$-enriched vertical excitation spectra of AB-bt bilayer silicene, e.g., form, intensity, number and frequency of special absorption structures.

\medskip
\par

Magnetic quantization phenomena of low-dimensional systems could be investigated using magneto-optical spectroscopies \cite{JZ;PRL2007, PP;PRL2008, PP;PRB2012, NRJ;PRL2013, OM;PRB2011}. The magnetic field is presented by a superconducting magnet \cite{JZ;PRL2007, PP;PRL2008, OM;PRB2011} and semidestructive singleturn coil \cite{PP;PRB2012, NRJ;PRL2013} with the desired field strength below 80 T. The examined/verified phenomena are exclusive in graphene-related systems, such as dispersionless LLs in layered graphenes \cite{JZ;PRL2007, PP;PRL2008, OM;PRB2011} and quasi-one-dimensional Landau subbands in bulk graphites \cite{JZ;PRL2007, PP;PRL2008, OM;PRB2011}. A lot of pronounced delta-function-like absorption peaks are clearly revealed by the inter-LL excitations arising from massless and massive Dirac fermions, respectively, in monolayer \cite{JZ;PRL2007} and AB-stacked bilayer graphenes \cite{OM;PRB2011}. For absorption peak frequencies, the former and the latter obviously exhibit the square-root and linear $B_z$-dependences. Concerning inter-Landau-subband excitations in Bernal graphite, one could observe a strong dependence on the wave vector $k_z$, which characterizes both kinds of Dirac quasi-particles \cite{PP;PRB2012, NRJ;PRL2013}. The  rich and unique magneto-optical spectra in bilayer AB-bt silicene are worthy of further experimental examinations, covering diverse absorption structures, $B_z$- and $E_z$-created excitation channels/threshold frequency, many absorption peaks within a very narrow frequency range, and the absence/presence of specific selection rules. They could provide the rather useful information about the buckled structure, stacking configuration/interlayer hopping integrals, and significant layer-dependent SOCs.

\medskip
\par

\section{Concluding Remarks}
\label{sec:4}

In conclusion, the electronic and optical properties of AB-bt bilayer silicene under electric and magnetic fields are studied using the generalized tight-binding model.
The theoretical framework is suitable for the full exploration of the essential properties in many other emergent
2D materials. The rich and unique critical properties are identified from the special buckled structure, symmetric stacking configuration, complicated intralayer and interlayer atomic interactions, and significant layer-dependent SOCs.  They are in great contrast with those in layered graphenes \cite{CY;IOP2017}.
The theoretical predictions could be verified by three kinds of optical spectroscopies \cite{ZQ;NP2008, LM;PRB2008, MKF;PRL2010, LZQ;PRL2009, JZ;PRL2007, PP;PRL2008, MKF;PRL2009, MKF;PRL2008, PP;PRB2012, NRJ;PRL2013, OM;PRB2011}.

\medskip
\par
AB-bt bilayer silicene possesses the feature-rich low-lying energy bands, with a sizable indirect band gap, the extremal band-edge states, the partially flat energy dispersions along the specific direction, and the constant-energy loops.
These critical points in energy-wave-vector space are revealed as shoulder-like structures and prominent peaks (van Hove singularities) in the DOS.
The absorption spectrum presents two shoulder structures, coming from band-edge states, and an antisymmetric peak, corresponding to the weak energy dispersions. An applied electric field can create the spin-split states, the semiconductor-metal phase transition and the formation of Dirac-cone structures at different valleys. As a result, more abnormal optical structures come to exist in the absorption spectra.
\medskip
\par

The magnetic quantization presents the unusual and complex phenomena.
The low-lying magneto-electronic structures cover four sublattice- and spin-dominated subgroups of LLs, in which they have the similar $B_z$ dependent energy spectrum without any crossing/anti-crossing behavior.
The conduction and valence LLs are doubly degenerate, being quantized from the different valley and thus creating the distinct localization centers.
There exists the oscillation modes overlapping for the LLs originated from the $\bf{K}$ and $\bf{T}$ points, while this is absent in graphene systems.
Such phenomenon enables the vertical optical transitions of LLs from different localization centers.
The magneto-absorption spectra consist of 4 intra-group and 12 inter-group inter-LL optical transition categories.
They are diversified by many single, double and twin non-uniform delta-function-like peaks under the absence of a specific selection rule. A lot of magneto-absorption peaks in a narrow frequency range are never observed in other 2D materials \cite{CY;IOP2017, ZQ;NP2008, LM;PRB2008, MKF;PRL2010, LZQ;PRL2009, JZ;PRL2007, PP;PRL2008, MKF;PRL2009, MKF;PRL2008, PP;PRB2012, NRJ;PRL2013, OM;PRB2011}.

\medskip
\par
The uniform perpendicular electric field leads to the dramatic changes in LLs, and thus the more complex magneto-absorption spectra. Since the oscillating band structure becomes pronounced, both well-behaved conduction and valence LLs come to exist at each valley. The splitting of sublattice- and localization-dominated LLs is getting more obvious.
Accordingly, the inter-LL optical transitions at lower frequency with specific selection rules are revealed in the absorption spectra. The threshold frequency gradually declines with the increasing $E_z$. The electric-field-controlled optical properties open a new opportunity in the application of novel designs of $Si$-based nano-electronics and optical-devices with enhanced mobilities.

\medskip
\par

\par\noindent {\bf Acknowledgments}

\medskip
\par
This material is based upon work supported by the Air Force Office of Scientific Research under award number FA2386-18-1-0120. We would also like to acknowledge the financial support from the Ministry of Science and Technology of the Republic of China (Taiwan) under Grant No. 105-2112-M-017-002-MY2. \\

\end{document}